\documentclass[11pt]{article}

\usepackage{latexsym}
\usepackage{amsmath}
\usepackage{float}
\usepackage{bm}
\usepackage{amssymb}
\usepackage{xcolor}
\usepackage{amsthm}
\usepackage{dsfont}
\usepackage[ansinew]{inputenc}
\usepackage{graphicx}
\usepackage{epsfig}
\usepackage{dsfont}
\usepackage{bbm}
\usepackage{url}
\usepackage{placeins}
\usepackage{multirow}
\usepackage{algorithm} 
\usepackage{enumitem}
\usepackage{overpic}

\makeatletter
\newenvironment{breakablealgorithm}
  {
   \begin{center}
     \refstepcounter{algorithm}
     \hrule height.8pt depth0pt \kern2pt
     \renewcommand{\caption}[2][\relax]{
       {\raggedright\textbf{\fname@algorithm~\thealgorithm} ##2\par}%
       \ifx\relax##1\relax 
         \addcontentsline{loa}{algorithm}{\protect\numberline{\thealgorithm}##2}%
       \else 
         \addcontentsline{loa}{algorithm}{\protect\numberline{\thealgorithm}##1}%
       \fi
       \kern2pt\hrule\kern2pt
     }
  }{
     \kern2pt\hrule\relax
   \end{center}
  }
\makeatother

\usepackage[colorlinks]{hyperref}
\hypersetup{allcolors =blue}

\usepackage[authoryear]{natbib}
\renewcommand{\epsilon}{\varepsilon}
\newcommand{\RR}{\mathds{R}}

\newcommand{\Prob}{\mathds{P}}
\newcommand{\Var}{\text{Var}}
\newcommand{\se}{\text{se}}
\newcommand{\argmax}{\operatornamewithlimits{argmax}}
\begin{document}

\title{{\bf \Large Overcoming model uncertainty -- how equivalence tests can benefit from model averaging
}}
\author{Niklas Hagemann$^{1}$ and
        Kathrin M\"ollenhoff$^{1}$  \\
 \small{$^{1}$ Institute of Medical Statistics and Computational Biology (IMSB),}  \\ \small Faculty of Medicine, University of Cologne, Germany 
 }

\maketitle

\begin{abstract}
    A common problem in numerous research areas, particularly in clinical trials, is to test whether the effect of an explanatory variable on an outcome variable is equivalent across different groups. In practice, these tests are frequently used to compare the effect between patient groups, e.g. based on gender, age or treatments. Equivalence is usually assessed by testing whether the difference between the groups does not exceed a pre-specified equivalence threshold. 
    Classical approaches are based on testing the equivalence of single quantities, e.g. the mean, the area under the curve (AUC) or other values of interest. However, when differences depending on a particular covariate are observed, these approaches can turn out to be not very accurate. Instead, whole regression curves over the entire covariate range, describing for instance the time window or a dose range, are considered and tests are based on a suitable distance measure of two such curves, as, for example, the maximum absolute distance between them.
    In this regard, a key assumption is that the true underlying regression models are known, which is rarely the case in practice. However, misspecification can lead to severe problems as inflated type I errors or, on the other hand, conservative test procedures.

    In this paper, we propose a solution to this problem by introducing a flexible extension of such an equivalence test using model averaging in order to overcome this assumption and making the test applicable under model uncertainty. Precisely, we introduce model averaging based on smooth AIC weights and we propose a testing procedure which makes use of the duality between confidence intervals and hypothesis testing. We demonstrate the validity of our approach by means of a simulation study and demonstrate the practical relevance of the approach considering a time-response case study with toxicological gene expression data.
\end{abstract}

\section{Introduction} \label{sec:intro}
In numerous research areas, particularly in clinical trials, a common problem is to test whether the effect of an explanatory variable on an outcome variable is equivalent across different groups \citep[see, e.g.,][]{Otto2008,Jhee2004}. 
Equivalence is usually assessed by testing whether the difference between the groups does not exceed a pre-specified equivalence threshold.
The choice of this threshold is crucial as it resembles the maximal amount of deviation for which equivalence can still be concluded. One usually chooses the threshold based on prior knowledge, as a percentile of the range of the outcome variable or resulting from regulatory guidelines.
Equivalence tests provide a flexible tool for plenty of research questions. For instance, they can be used to test for equivalence across patient groups, e.g. based on gender or age, or between treatments.
Moreover, they are a key ingredient of bioequivalence studies, investigating whether two formulations of a drug have nearly the same effect and are hence considered to be interchangeable \citep[e.g.][]{moellenhoff2022, Hauschke2007}. 

Classical approaches are based on testing the equivalence of single quantities, e.g. the mean, the area under the curve (AUC) or other values of interest \citep{Schuirmann1987, Lakens2017}. However, when differences depending on a particular covariate are observed, these approaches can turn out to be not very accurate. Instead, considering the entire covariate range, describing for instance the time window or a dose range, has recently been proposed by testing equivalence of whole regression curves. Such test are typically based on the principle of confidence interval inclusion \citep{Liu2009, Gsteiger2011, Bretz2018}. However, a more direct approach applying various distance measures has been introduced by \citet{Dette2018}, which turned out to be particularly more powerful. Based on this, many further developments, e.g. for different outcome distributions, specific model structures and/or responses of higher dimensions, have been introduced \citep[see, e.g.,][]{Moellenhoff2020, Moellenhoff2021, Moellenhoff2024, Hagemann2024}.

All these approaches have one thing in common: they base on the assumption that the true underlying regression model is known. In practice this usually implies that the models need to be chosen manually, either based on prior knowledge or visually. 
Hence, these approaches might not be robust with regard to model misspecification and, consequently, suffer from problems like inflated type I errors or reduced power \citep{Guhl2022, Dennis2019}.
One idea to tackle this problem is implementing a testing procedure which explicitly incorporates the model uncertainty. This can be based on a formal model selection procedure, see, e.g., \citet{Moellenhoff2018}, who propose conducting a classical model choice procedure prior to preforming the equivalence test. 

An alternative to this is the incorporation of a model averaging approach into the test procedure.
As outlined by \cite{Bornkamp2015} model selection has some disadvantages compared to model averaging. Particularly, model selection is not stable in the sense that minor changes in the data can lead to major changes in the results \citep{Breiman1996}. This also implies that model selection is non-robust with regard to outliers. In addition, the estimation of the distribution of post model selection parameter estimators is usually biased 
\citep{Leeb2005, Leeb2008}.
Model averaging is omnipresent whenever model uncertainty is present, which is, besides other applications, often the case in parametric dose response analysis. Besides practical applications, there are also several methodological studies regarding model averaging in dose-response studies \citep[see, e.g.,][]{Schorning2016, Aoki2017, Buatois2018} and \citet{Bornkamp2009}, who incorporated model averaging as an alternative to model selection in their widely used dose-finding method MCPMod.

Therefore, in this paper, we propose an approach utilising model averaging rather than model selection. 
There are frequentist as well as Bayesian model averaging approaches. 
The former almost always use the smooth weights structure introduced by \citet{Buckland1997}. These weights depend on the values of an information criterion of the fitted models. Predominantly, the Akaike information criterion \citep[AIC;][]{AIC} is used but other information criteria can be used as well.
While only few of the Bayesian approaches perform fully Bayesian inference \citep[see, e.g.,][]{Ley2009}, the majority makes use of the fact that the posterior model probabilities can be approximated by weights based on the Bayesian information criterion \citep[BIC;][]{BIC} that have the same smooth weights structure as the frequentist weights \citep{Wasserman2000}. 

In this paper, we propose an equivalence test incorporating model-averaging and hence overcoming the problems caused by model uncertainty. Precisely, we first make use of the duality between confidence intervals and hypothesis testing and propose a test based on the derivation of a confidence interval. By doing so, we both guarantee numerical stability of the procedure and provide confidence intervals for the measure of interest.
We demonstrate the usefulness of our method with the example of toxicological gene expression data. 
In this application, using model averaging enables us to analyse the equivalence of time-response curves between two groups for 1000 genes of interest without the necessity of specifying all 2000 correct models separately, thus avoiding both a time-consuming model selection step and potential model misspecifications.

The paper is structured as follows: 
In Section \ref{sec:MA}, dose-response models and the concept of model averaging are succinctly discussed. 
In Section \ref{sec:EquiTest}, the testing approach is introduced, proposing three different variations. Finite sample properties in terms of Type I and II error rates are studied in Section \ref{sec:simu}.
Section \ref{sec:case} illustrates the method using the toxicological gene expression example before Section \ref{sec:conclusion} closes with a discussion.

\section{Model averaging for dose-response models} \label{sec:MA}
\subsection{Dose-response models} \label{sec:MA:regression}
We consider two different groups, indicated by an index $l = 1, 2$, with corresponding response variables $y_{lij}$ 
with $\mathcal{Y} \subseteq \RR$ denoting the set of all possible outcomes. There are $i = 1, ..., I_l$ dose levels and $j = 1, ..., n_{li}$ denotes the observation index within each dose level. For each group the total number of observations is $n_l$ and $n$ is the overall number of observations, i.e. $n_l = \sum_{i=1}^{I_l} n_{li}$ and $n=n_1 + n_2$.
For each group we introduce a flexible dose-response model  
$$y_{lij}=m_l(x_{li}, \boldsymbol{\theta}_{l}) + e_{lij}, \quad j = 1, ..., n_{li}, \quad i = 1, ..., I_l, \quad l=1,2,$$
where 
$x_{li} \in \mathcal{X} \subseteq \RR$ is the dose level, i.e. the deterministic explanatory variable. We assume the 
residuals $e_{lij}$ to be independent, have expectation zero and finite variance $\sigma^2_l$.
The function $m_l(\cdot)$ models the effect of $x_{li}$ on $y_{lij}$ via a regression curve with $\boldsymbol{\theta}_{l} \in \RR^{\dim(\boldsymbol{\theta}_{l})}$ being its parameter vector. 
We assume $m_{l}(\cdot)$ to be twice continuously differentiable.
In dose-response studies, as well as in time-response studies, often either a linear model
\begin{equation} \label{eq:linear}
    m_l(x, \theta_l) = \beta_{l0} + \beta_{l1} x,
\end{equation}
a quadratic model
\begin{equation} \label{eq:quadratic}
    m_l(x, \theta_l) = \beta_{l0} + \beta_{l1} x + \beta_{l2} x^2,
\end{equation}
an emax model
\begin{equation} \label{eq:emax}
    m_l(x, \theta_l) = \beta_{l0} + \beta_{l1} \frac{x}{\beta_{l2} + x},
\end{equation}
an exponential (exp) model 
\begin{equation} \label{eq:exp}
    m_l(x, \theta_l) = \beta_{l0} + \beta_{l1} \left( \exp \left( \frac{x}{\beta_{l2}} \right) -1 \right),
\end{equation} 
a sigmoid emax (sigEmax) model
\begin{equation} \label{eq:sigemax}
    m_l(x, \theta_l) = \beta_{l0} + \beta_{l1} \frac{x^{\beta_{l3}}}{\left( \beta_{l2}\right)^{\beta_{l3}} + x^{\beta_{l3}}},
\end{equation}
also known as Hill model or 4pLL-model, or a beta model
\begin{equation} \label{eq:beta}
    m_l(x, \theta_l) = \beta_{l0} + \beta_{l1} \left(\frac{\left( \beta_{l2} + \beta_{l3}\right)^{\beta_{l2} + \beta_{l3}}}{\left( \beta_{l2}\right)^{\beta_{l2}}+ \left( \beta_{l3}\right)^{\beta_{l3}}}\right) \left(\frac{x}{s}\right)^{\beta_{l2}}  \left(1 - \frac{x}{s} \right) ^{\beta_{l3}},
\end{equation}
where $s$ is a fixed scaling parameter, is deployed 
\citep[][]{Bretz2005, Pinheiro2006, Pinheiro2014, Duda2022}.
These models strongly vary in the assumed underlying dose-response relation, e.g. in terms of monotony, and consequently in the shape of their curves. Therefore, choosing a suitable dose-response model is crucial for all subsequent analyses.  

However, in practical applications the true underlying model shape is in general unknown. Thus, it might not always be clear which functional form of \ref{eq:linear}-\ref{eq:beta} should be imployed. A possible answer to this is implementing
model averaging which, as outlined in Section \ref{sec:intro}, has several advantages over the simpler alternative of model selection. 

\subsection{Model averaging} \label{sec:MA:MA}
As outlined before, frequentist as well as Bayesian model averaging approaches usually both use the same smooth weights structure introduced by \citet{Buckland1997} and \citet{Wasserman2000}, respectively. 
Accordingly, by leaving out the group index $l=1,2$ for better readability 
the averaged model is given by 
\begin{equation} \label{eq:MA}
    m(x, \hat{\boldsymbol{\theta}}) := \sum_{k=1}^K w_{k} \, m_{k}(x, \hat{\boldsymbol{\theta}}_k),
\end{equation}
where the $m_k(x, \hat{\boldsymbol{\theta}}_{k}), k =1,..,K,$ correspond to the  $K$ candidate models,
\begin{equation} \label{eq:weights}
    w_{k} = \frac{\exp(0.5 \, I(m_k(x, \hat{\boldsymbol{\theta}}_{k}))}{\sum_{\Tilde{k}=1}^K \exp(0.5 \, I( m_{\Tilde{k}}(x, \hat{\boldsymbol{\theta}}_{\Tilde{k}})))}
\end{equation}
%
%
are the corresponding weights and $I(\cdot)$ is an information criterion. Usually the AIC is used for frequentist model averaging, while the BIC is usually deployed for Bayesian model averaging \citep{Schorning2016}.  

Despite the prevalence of the AIC and BIC, other information criteria are sometimes used as well, e.g. the deviance information criterion 
\citep[see, e.g.,][]{Price2011}.
Occasionally, model averaging also bases on cross-validation or machine learning methods. 
For a more general introduction to model averaging techniques the reader is refereed to e.g. \citet{Fletcher_2018} or \citet{Claeskens_2008} and an overview specifically focusing on dose-response models is given by \citet{Schorning2016}.  
\subsection{Inference} \label{sec:MA:inference}
As the parameter estimation is conducted for each of the candidate models separately, it is not influenced by the subsequent model averaging. Therefore, here the index $k$ is left out. 
Inference can be based on an ordinary least squares (OLS) estimator, i.e. minimising 
$$
\sum_{i=1}^{I_l} \sum_{j=1}^{n_{li}} (y_{lij} - m_{l}(x_{li}, \boldsymbol{\theta}_{l}))^2, \quad l=1,2.
$$ 
Alternatively, under the assumption of normality of the residuals, i.e.
$$e_{lij} \stackrel{iid}{\sim} N(0, \sigma^2_{l}), \quad l=1,2,$$
a maximum likelihood estimator (MLE) can also be deployed where the log-likelihood is given by
\begin{equation} \label{eq:logLik}
   \ell(\boldsymbol{\theta}_{l}, \sigma_l^2 
   ) = -\frac{n_l}{2} \ln(2\pi\sigma_l^2) - \frac{1}{2\sigma_l^2} \sum_{i=1}^{I_l} \sum_{j=1}^{n_{li}} (y_{lij} - m_{l}(x_{li}, \boldsymbol{\theta}_{l}))^2, \quad l=1,2. 
\end{equation}
Under normality both approaches are identical and, hence, lead to the same parameter estimates $\hat{\boldsymbol{\theta}}_{l}$. From \eqref{eq:logLik} an MLE for the variance 
\begin{equation} \label{eq:var}
\hat{\sigma}_l^2 = \frac{1}{n_l} \sum_{i=1}^{I_l} \sum_{j=1}^{n_{li}} (y_{lij} - m_{l}(x_{li}, \boldsymbol{\theta}_{l}))^2, \quad l=1,2
\end{equation}
can be derived as well. 
In \texttt{R} inference is performed with the function \texttt{fitMod} from the package \texttt{Dosefinding} \citep{Bornkamp2009, Pinheiro2014} which performs OLS estimation. The value of the log-likelihood needed for the AIC or BIC is then calculated by plugging the OLS estimator into the log-likelihood \eqref{eq:logLik}. 

\section{Model-based equivalence tests under model uncertainty} \label{sec:EquiTest}

\subsection{Equivalence testing based on confidence intervals} \label{sec:EquiTest:CI}
Model-based equivalence tests have been introduced in terms of the $L^2$-distance, the $L^1$-distance or the maximal absolute deviation (also called $L^\infty$-distance) of the model curves \citep{Dette2018, Bastian2024}. 
Although all of these approaches have their specific advantages and disadvantages as well as specific applications, subsequent research \citep[see, e.g.,][]{Moellenhoff2018, Moellenhoff2020, Moellenhoff2021, Hagemann2024} is predominately based on the maximal absolute deviation due to its easy interpretability.
Accordingly, we state the hypotheses 
\begin{equation} \label{eq:hypotheses}
    H_0: d \geq \epsilon \text{   vs.   } H_1: d < \epsilon
\end{equation}
of equivalence of regression curves with respect to 
the maximal absolute deviation,
that is
$$
d = \max_{x \in \mathcal{X}}|m_1(x, \boldsymbol{\theta}_{1}) - m_2(x, \boldsymbol{\theta}_{2})|
$$
is the maximal absolute deviation of the curves and $\epsilon$ is the pre-specified equivalence threshold, i.e. that a difference of $\epsilon$ is believed
not to be clinically relevant.
The test statistic is given as the estimated maximal deviation between the curves
\begin{equation} \label{eq:test_stat}
    \hat{d} = \max_{x \in \mathcal{X}}|m_1(x, \hat{\boldsymbol{\theta}}_{1}) - m_2(x, \hat{\boldsymbol{\theta}}_{2})|.
\end{equation}

 As the distribution of $\hat d$ under the null hypothesis is in general unknown, it is usually either approximated based on a parametric bootstrap procedure or by asymptotic theory.
In \citet{Dette2018} the asymptotic validity of both approaches is proven, but the corresponding simulation study shows that the bootstrap test outperforms the asymptotic test in finite samples. For the bootstrap test several studies \citep[see, e.g.,][]{Dette2018, Moellenhoff2018, Moellenhoff2020} show reasonable results for finite samples across applications. 

However, in light of practical application, this approach can have two disadvantages: First, it does not directly provide confidence intervals (CI) which provide useful information about the precision of the test statistic. 
Further, they would have an important interpretation analogously to their interpretation in classical equivalence testing known as TOST \citep[two one-sided tests;][]{Schuirmann1987}, where the bounds of the confidence interval are typically compared to the confidence region of $[-\epsilon,\epsilon]$. 

Second, it requires the estimation of the models under the constraint of being on the edge of the null hypothesis, i.e. the maximal absolute deviation being equal to $\epsilon$ \citep[see Algorithm 1 in][]{Dette2018}.
Technically, this is usually conducted using augmented Lagrangian optimisation.
However, with increasing model complexity, this becomes numerically challenging. In the context of model averaging, these numerical issues are particularly relevant since all models would need to be estimated jointly as they need to jointly fulfil the constraint. This leads to a potentially high dimensional optimisation problem with a large number of parameters. In addition, for model averaging the side constraint has a highly complex structure because with every parameter update not only the model curves change but also the model weights do.

As an alternative to approximating the distribution under the null hypothesis, we propose to test hypotheses \eqref{eq:hypotheses} based on the well-known duality between confidence intervals and hypothesis testing \citep{Aitchison1964}. This testing approach is similar to what \citet[][Algorithm 1]{Bastian2024} introduced for the L$^1$ distance of regression models. 
Therefore, let $(-\infty, u]$ be a one-sided lower $(1-\alpha)$-CI for $d$ which we can rewrite as $[0, u]$ due to the non-negativity of $d$, i.e.
$$
\Prob(d \leq u)=\Prob(d \in (-\infty, u])=\Prob(d \in [0, u]) \geq 1-\alpha.
$$

According to the duality between CI and hypothesis testing, we reject the null hypothesis and conclude equivalence if
\begin{equation} \label{eq:test_decision}
    \epsilon > u.
\end{equation}
This testing procedure is an $\alpha$-level test as
\begin{align*}
    & \, \Prob_{H_0}(\epsilon > u)\\
    \leq & \, \Prob(d > u) \\
    =& \, 1 - \Prob(d \leq u) \\
    \leq & \, 1-(1-\alpha) = \alpha.
\end{align*}

However, as the distribution of $\hat d$ is in general unknown, obtaining $u$ is again a challenging problem. It is obvious from \eqref{eq:test_decision} that the quality of the testing procedure crucially depends on the quality of the estimator for $u$. If the CI is too wide
the test procedure is conservative and lacks power. In contrast, a too narrow CI can lead to type I
error inflation due to not reaching the desired coverage probability $1- \alpha$.
We propose three different possibilities to calculate the CI, namely
\begin{enumerate}
    \item CI based on a parametric percentile bootstrap,
    \item asymptotic CI based on the asymptotic distribution of $\hat d$ derived by \citet{Dette2018}, and
    \item a hybrid approach using the asymptotic normality of $\hat d$ but estimating its standard error based on a parametric bootstrap.
\end{enumerate}

One-sided CIs based on a parametric percentile bootstrap can be constructed in the same way \citet{Moellenhoff2018} proposed for two-sided CIs. 
In order to do so, they obtain parameter estimates (either via OLS or maximum likelihood optimisation), generate bootstrap data from these estimates and calculate  the percentiles from the ordered bootstrap sample.
The resulting test is similar to what \citet[][Algorithm 1]{Bastian2024} derived for the L$^1$ distance of regression models. That is
$$
\left[0, \hat{q}^*(1- \alpha)\right],
$$
where $\hat{q}^*(1- \alpha)$ denotes the $(1-\alpha)$-quantile of the ordered bootstrap sample. 

Asymptotic CIs can be derived directly from test (5.4) of \citet{Dette2018} and are given by 
\begin{equation} \label{eq:CIasymp}
\left[0, \hat{d} + \sqrt{\frac{\widehat{\Var}(d)}{n}} z \right]
\end{equation}
where $z$ is the $(1-\alpha)$-quantile of the standard normal distribution and $\widehat{\Var}(d)$ is the closed-form estimator for the variance of $d$ given by equation (4.7) of \citet{Dette2018}. 
However, the asymptotic validity of this variance estimator is only given under the assumption that within $\mathcal{X}$ there is only one unique value $x_0$ 
where the absolute difference curve attains its maximum, i.e.
$x_0 = \argmax_{x \in \mathcal{X}}|m_1(x, \boldsymbol{\theta}_{1}) - m_2(x, \boldsymbol{\theta}_{2})|$ 
 and, moreover, that this value $x_0$ is known. 
This does not hold in general as \citet{Dette2018} give two explicit counterexamples in terms of two shifted emax or exponential models. In addition, in practical applications $x_0$ is in general not known and needs to be estimated. If the absolute deviation along $x$ is small, the estimation of $x_0$ can become unstable leading to an unstable variance estimator. 
Moreover, as mentioned before, the simulation study of \citet{Dette2018} shows that for finite samples the bootstrap test is superior to the asymptotic test. 

Given the disadvantages of the asymptotic CI and especially of the underlying variance estimator, we introduce a hybrid approach which is a combination of both approaches. It is based on the asymptotic normality of $\hat d$ but estimates the standard error of $\hat{d}$ based on a parametric bootstrap 
leading to 
$$
\left[0, \hat{d} + \widehat{\se}(\hat{d}) z \right],
$$
where the estimator $\widehat{\se}(\hat{d})$ of the standard error of $\hat{d}$ is the empirical standard deviation of the bootstrap sample. 

Under the assumptions introduced by \citet{Dette2018}, all three approaches are asymptotically valid. For the test based on the asymptotic CI, this follows directly from \citet{Dette2018}. This also applies to the hybrid CI-based test due to $\widehat{\se}(\hat{d})$ being an asymptotically unbiased estimator for the standard error of $\hat{d}$ as outlined by \citet{Efron1994}.
The asymptotic validity of the 
percentile approach follows from \citet[][Appendix: proof of Theorem 4]{Dette2018} analogously to how \citet[][Appendix A.3]{Bastian2024} derived the validity of their approach from there.
The finite sample properties of the three methods are compared in Section \ref{sec:simu:CI}.

\subsection{Model-based equivalence tests incorporating model averaging} \label{sec:EquiTest:MA}
We now combine the model averaging approach presented in Section \ref{sec:MA:MA} with the CI-based test introduced in Section \ref{sec:EquiTest:CI}. For the asymptotic test that is estimating $m_l(x, \hat{\boldsymbol{\theta}}_{l}), \, l=1,2$ using \eqref{eq:MA}
with model weights \eqref{eq:weights} and then calculating the test statistic \eqref{eq:test_stat}. 
Subsequently, the asymptotic CI \eqref{eq:CIasymp} can be determined using the closed form variance estimator given by \citet{Dette2018}. Using this CI, the test decision is based on decision rule \eqref{eq:test_decision}.

The testing procedure of the percentile and hybrid approach is shown in Algorithm \ref{alg1}, where the first two steps are essentially the same as for the asymptotic test. The percentile test is conducted by performing Algorithm step 4a, while conducting step 4b instead leads to the hybrid test. In the following we will refer to this as Algorithm \ref{alg1}a and Algorithm \ref{alg1}b, respectively.

\begin{breakablealgorithm} \caption{} \label{alg1} 

  \begin{enumerate}
      \item Obtain parameter estimates $\hat{\boldsymbol{\theta}}_{lk}, \, k=1,...,K_l, \, l = 1,2$ for the candidate models, either via OLS or maximum likelihood optimisation (see Section \ref{sec:MA:inference}). 
      Determine the averaged models from the candidate models using \eqref{eq:MA}, i.e. by calculating
      $$
      m_l(x, \hat{\boldsymbol{\theta}_l}) = \sum_{k=1}^{K_l} w_{lk} \, m_{lk}(x, \hat{\boldsymbol{\theta}}_{lk}), \quad l=1,2,
      $$ 
      with weights \eqref{eq:weights} as well as the variance estimator $\hat{\sigma}_l^2, \, l=1,2$ from \eqref{eq:var}.
      \item  Calculate the test statistic \eqref{eq:test_stat}.
	\item Execute the following steps:
			\begin{enumerate}[label=3.\arabic*]
				\item Obtain bootstrap samples by generating data according to the model parameters $\hat{\boldsymbol{\theta}}_{l}, \, l = 1,2$. Under the assumption of normality that is
    $$
    y^*_{lij} \sim N(m_l(x_{li}, \hat{\boldsymbol{\theta}}_{l}), \hat{\sigma}_l^2), \quad i=1,...,n_l, \, l=1,2.
    $$
				\item From the bootstrap samples, estimate the models  $m_{l}(x_{li}, \hat{\boldsymbol{\theta}}^{*}_{l}), \, l = 1, 2$ as in step $(1)$ and the test statistic
				\begin{equation}
				    \label{boot}
        \hat{d}^* = \max_{x \in \mathcal{X}}|m_{1}(x, \hat{\boldsymbol{\theta}}_{1}^{*}) - m_{2}(x, \hat{\boldsymbol{\theta}}_{2}^{*})|.
        \end{equation}
		\item Repeat steps (3.1) and (3.2) $n_{boot}$ times to generate replicates $\hat{d}^*_{1}, \dots, \hat{d}^*_{n_{boot}}$ of $\hat{d}^*$.
            Let $\hat{d}^*_{(1)} \leq \ldots \leq \hat{d}^*_{(n_{boot})}$
			denote the corresponding order statistic. 
            \end{enumerate}	
        \item Calculate the CI using one of the following approaches: 
\begin{enumerate}[label=\alph*]
				\item \underline{Percentile CI}: Obtain the estimated right bound of the percentile bootstrap CI as the $(1-\alpha)$-quantile of the bootstrap sample $$
\hat{u} = \hat{q}^*(1- \alpha)= \hat{d}^*_{(\lfloor n_{boot} (1-\alpha) \rfloor)}.
$$
    \item \underline{Hybrid CI}: Obtain the estimator for the standard error of $\hat{d}$ as $\widehat{\se}(\hat{d})= \sqrt{\widehat{\Var}(\hat{d}^*_1,...,\hat{d}^*_{n_{boot}})}$ and the estimated right bound of the hybrid CI as
    $$\hat{u} = \hat{d} + \widehat{\se}(\hat{d}) z.$$
            \end{enumerate}	

        \item 
            Reject the null hypothesis in \eqref{eq:hypotheses} and assess equivalence if $$\epsilon > \hat{u}.$$
	\end{enumerate}		

\end{breakablealgorithm}
\section{Finite sample properties} \label{sec:simu}
In the following we investigate the finite sample properties of the proposed tests by a simulation study. In order to ensure comparability, we reanalyse the simulation scenarios given by \cite{Dette2018}.
The dose range is given by $\mathcal{X}=[0,4]$ and data is observed for dose levels $x=0, 1,2,3$ and 4 with equal number of observations $n_{li} = \frac{n_l}{5}$ for each dose level.
All three simulation scenarios use the same three variance configurations $(\sigma_1^2, \sigma_2^2) \in \{(0.25, 0.25), (0.25, 0.5), (0.5, 0.5)\}$ as well as the same four different sample sizes $(n_1, n_2) \in \{(10, 10), (10, 20), (20, 20), (50, 50) \}$ and the same significance level of $\alpha = 0.05$.

In the first simulation scenario the equivalence of an emax model and an exponential model is investigated. The other two simulation scenarios consist of testing for equivalence of two shifted models, either both being emax models or both being exponential models. 
In contrast to the first scenario where the absolute deviation of the models is observed at one unique $x_0$, this is not the case for the latter two scenarios. 
Here, the deviation of both models is constant across the whole dose range $\mathcal{X}=[0,4]$, i.e. 
$$
|m_1(x, \boldsymbol{\theta}_{1}) - m_2(x, \boldsymbol{\theta}_{2})| = d \, \, \forall \, \, x \in \mathcal{X}
$$
as $m_1$, $m_2$ are just shifted. 

Hence, for these two scenarios the asymptotic test is not applicable as its close form variance estimator bases on the uniqueness of $x_0$. Therefore, only simulation scenario 1 is used to compare the three CI-based tests to each other as well as to the results observed by \citet{Dette2018}. 
Subsequently, all three scenarios are used to compare the performance of the test using model averaging to the one based on the correct specification of the models as well as under model misspecification. 

\subsection{Finite sample properties of confidence interval-based equivalence testing} \label{sec:simu:CI}
Prior to the investigation of the effect of model averaging onto the finite sample properties, we first inspect the performance of the CI-based testing approach, i.e. test \eqref{eq:test_decision}, itself. 
For the asymptotic test, the CI is defined by \eqref{eq:CIasymp}. 
For the percentile as well as the hybrid approach, the tests are conducted as explained in Section \ref{sec:EquiTest:CI} which is formally defined by setting $K_1 = K_2 = 1$ in Algorithm \ref{alg1}. 

As outlined before, simulation scenario 1 of \citet{Dette2018} is given by tesing for the equivalence of an emax model \eqref{eq:emax}
with $\theta_1 = (\beta_{10}, \beta_{11}, \beta_{12}) = (1, 2, 1)$ and an exponential model \eqref{eq:exp}
with $\theta_2 = (\beta_{20}, \beta_{21}, \beta_{22}) = (\beta_{20}, 2.2, 8)$.
It consists of 60 sub-scenarios resulting from the three different variance configurations each being combined with the four different sample sizes and
five different choices of $\beta_{20} \in \{0.25, 0.5, 0.75, 1, 1.5\}$, leading to the corresponding deviations of the regression curves being $d \in \{1.5, 1.25, 1, 0.75, 0.5\}$.
The test is conducted for $\epsilon = 1$ such that the first three deviations are under the null hypothesis and, therefore, used to investigate the type I error rates. The latter two deviations correspond to the alternative and are used to estimate the power of the tests. 

\begin{figure}
        \centering
        \includegraphics[scale = 0.795]{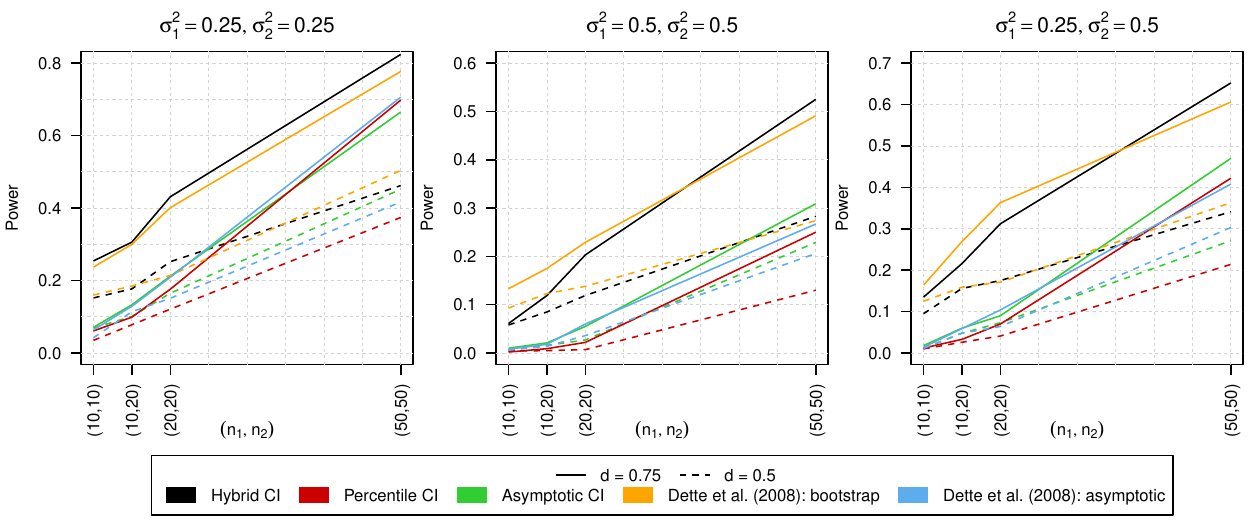} 
        \caption{Comparison of the power of the CI-based testing approaches to the testing approaches proposed by \citet{Dette2018} with $\epsilon = 1$. The results are shown for two distances of the regression curves $d \in \{0.5, 0.75\}$ and three different combinations of variances $(\sigma_1^2, \sigma_2^2) \in \{(0.25, 0.25), (0.5, 0.5), (0.25, 0.5)\}.$}
        \label{fig:CI_Power} 
\end{figure}

As the type I error rates are always smaller than the nominal level of $\alpha = 0.05$ for all three approaches (see table S1 of the supplementary material for exact values), i.e. all testing approaches always hold the nominal level, the following analysis focuses on the power of the tests. 
Figure \ref{fig:CI_Power} shows the power for all three tests for all sub-scenarios under the alternative as well as the corresponding power of the tests of \cite{Dette2018}. In each sub-scenario we observe that the hybrid test has superior power compared to the other two CI-based tests. In addition, one can observe that the power achieved by the hybrid test is quite similar to the one \citet{Dette2018} observed for their bootstrap test and, therefore, is also superior to the power of their asymptotic test. The power of the test based on the percentile CI is considerably smaller which indicates that the test might be overly conservative in finite samples. The test based on the asymptotic CI leads to nearly the same results as the asymptotic test of \citet{Dette2018} which is not surprising as it is directly derived from it. Consequentially, the lack of power that \citet{Dette2018} observed for their asymptotic test in comparison to their bootstrap test, is also present for the test based on the asymptotic CI. 

In conclusion, the hybrid approach which provides numerically advantages compared to the bootstrap test of \citet{Dette2018} and also leads to additional interpretability due to providing CIs, achieves nearly the same power as the bootstrap test while holding the nominal level. 

\subsection{Finite sample properties under model uncertainty} \label{sec:simu:MA}
We now investigate the finite sample properties under model uncertainty. Due to the clear superiority of the hybrid approach observed in Section \ref{sec:simu:CI}, only the hybrid test is used for this analysis. 
We compare the performance of the test using model averaging to the one based on the correct specification of the models as well as under model misspecification. 
We use frequentist model averaging with AIC-based weights as the inference is frequentist as well. The corresponding equivalence tests are conducted using Algorithm \ref{alg1}b.

\subsubsection{Comparison of an emax
with an exponential model} \label{sec:simu:MA:1}
\begin{figure}  
        \centering
        \includegraphics[scale = 0.795]{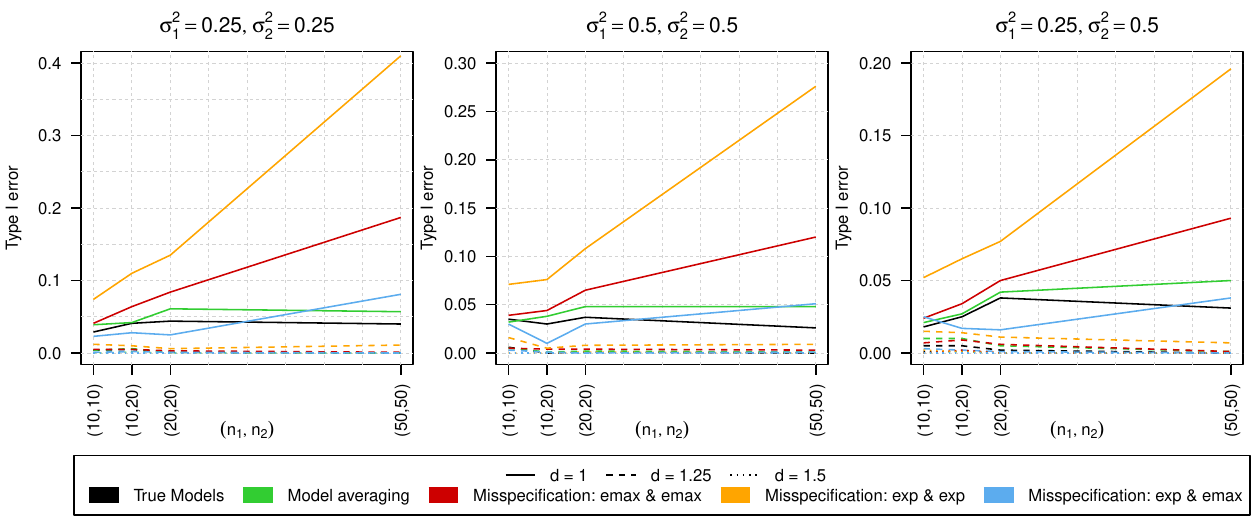}
        \caption{Comparison of the type I error rates of the test using the true model, the model averaging-based test and the tests under model misspecification in scenario 1. The results are shown for $\epsilon = 1$, three distances of the regression curves $d \in \{1, 1.25, 1.5\}$ and three different combinations of variances $(\sigma_1^2, \sigma_2^2) \in \{(0.25, 0.25), (0.25, 0.5), (0.5, 0.5)\}.$}
        \label{fig:Ex1_Type1}
\end{figure}

First, we again investigate the first simulation scenario introduced in Section \ref{sec:simu:CI} but now under model uncertainty where it is unclear if an emax or an exponential model applies for each of the groups implying $K_1 = K_2 = 2$ and leading to one correct specification, as well as three misspecifications. 
Figure \ref{fig:Ex1_Type1} shows the corresponding type I error rates for all sub-scenarios under the null hypothesis, i.e. for $d \in \{1, 1.25, 1.5\}$ and $(\sigma_1^2, \sigma_2^2) \in \{(0.25, 0.25), (0.25, 0.5), (0.5, 0.5)\}$ for the correct specification, the three misspecifications as well as under model averaging.  

One can observe that falsely specifying the same model for both responses leads to highly inflated type I error rates which, in addition, even increase for increasing sample size.
The highest type I errors are present if an exponential model is specified for both responses leading to type I error rates being as large as 0.410 which is observed for $\sigma_1^2  = \sigma_2^2 = 0.25$ and $n_1 = n_2 = 50$. 
If an emax model is specified for both responses, the type I error inflation is smaller but still present and reaches up to 0.187 which is observed for $\sigma_1^2  = \sigma_2^2 = 0.25$ and $n_1 = n_2 = 50$.
The third misspecification under investigation is that specifying the models the wrong way round, i.e. an exponential model for the first group and an emax model for the second one. In comparison to the other two misspecifications, this leads to less extreme results but type I error inflation is still observable. 
This results from the fact that using one convex (exponential) and one concave (emax) model usually leads to a larger  maximal absolute deviation than using two convex or two concave models and, therefore, in general to fewer rejections of the null hypothesis.

Compared to these results, the type I errors resulting from model averaging are closer to the nominal level of the test. However, for two out of the 36 investigated sub-scenarios ($\sigma_1^2  = \sigma_2^2 = 0.25$ and $n_1 = n_2 = 20$ as well as $n_1 = n_2 = 50$) the type I errors still exceeds the nominal level but to a much lesser extent compared to model misspecification, reaching a maximum of 0.061. As expected, when using the true underlying model, the test holds the nominal level of $\alpha = 0.05$.
Comparison of the power of the tests is not meaningful as some of them are not holding the nominal level. However, the estimated power is shown in table S2 of the supplementary material.  
\subsubsection{Comparison of two shifted emax models} \label{sec:simu:MA:2}
We continue by investigating the fine sample properties for the case of two shifted emax models, i.e. model \ref{eq:emax} now applies for both groups, where $\theta_1 = (\beta_{10}, \beta_{11}, \beta_{12}) = (\beta_{10}, 5, 1)$ and $\theta_2 = (\beta_{20}, \beta_{21}, \beta_{22}) = (0, 5, 1)$, which implies $d = \beta_{10}$. The levels of $d$ under investigation are $1, 0.75, 0.5, 0.25, 0.1$ and $0$. The test is conducted for $\epsilon = 0.5$ such that the first three deviations are under the null hypothesis and, therefore, used to investigate the type I error rates. 
The latter three deviations are under the alternative and used to estimate the power of the tests.  

\begin{figure}
        \centering
        \includegraphics[scale = 0.795]{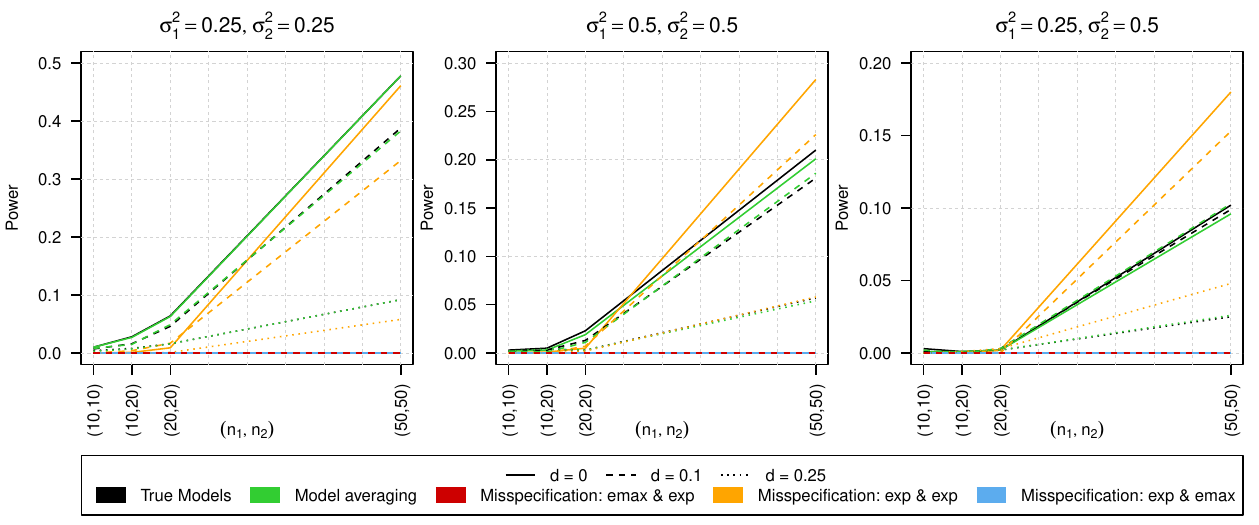}
        \caption{Comparison of the power of the test using the true model, the model averaging-based test and the tests under model misspecification in scenario 2. The results are shown for $\epsilon = 0.5$, three distances of the regression curves $d \in \{1, 1.25, 1.5\}$ and three different combinations of variances $(\sigma_1^2, \sigma_2^2) \in \{(0.25, 0.25), (0.25, 0.5), (0.5, 0.5)\}.$}
        \label{fig:Ex2_Power} 
\end{figure}

We only observe few type I error rates which are non-zero and these are still much smaller than the nominal level of $\alpha = 0.05$, reaching a maximum of only $0.003$ (all values can be found in table S3 of the supplementary material). Hence, the analysis focuses on the power of the tests which is shown in Figure \ref{fig:Ex2_Power}.
One can observe falsely specifying one of the models to be an exponential model leads to the power being constantly equal to zero even for sub-scenarios which are quite far under the alternative. For misspecification in terms of using an exponential model for both responses, the power loss is not that extensive but still occurs for smaller sample sizes, which is especially visible for $\sigma^2_1 = \sigma^2_2 = 0.25$ due to the estimation uncertainty being  the smallest. 
In contrast, model averaging results in nearly the same power as using the true model. This also leads to the fact that in some cases the black line is even hardly visual as it is nearly perfectly overlapped by the green one.

\subsubsection{Comparison of two shifted exponential models} \label{sec:simu:MA:3}
The third simulation scenario is given by two shifted exponential models, i.e. model \ref{eq:exp} now applies for both groups, where $\theta_1 = (\beta_{10}, \beta_{11}, \beta_{12}) = (\beta_{10}, 2.2, 8)$ and $\theta_2 = (\beta_{20}, \beta_{21}, \beta_{22}) = (0, 2.2, 8)$ which implies $d = \beta_{10}$, resulting in the same values for $d$ as in Section \ref{sec:simu:MA:2}. The test is conducted for $\epsilon = 0.5$ such that the first three deviations are under the null hypothesis and, therefore, used to investigate the type I error rates. The latter three deviations are under the alternative and used to estimate the power of the tests.  

\begin{figure} 
        \centering
        \includegraphics[scale = 0.79]{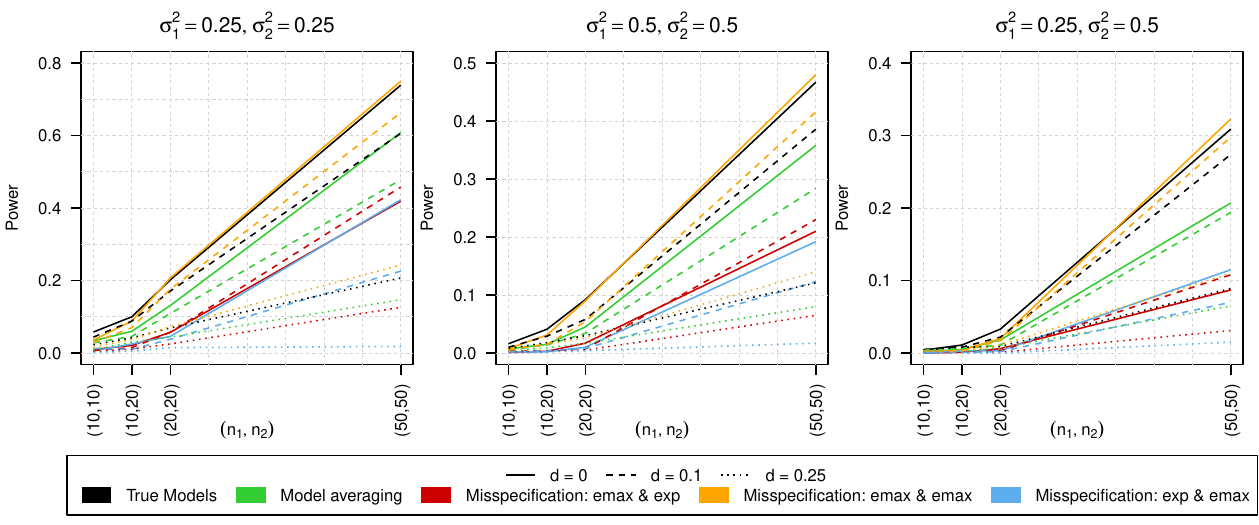}
        \caption{Comparison of the power of the test using the true model, the model averaging-based test and the tests under model misspecification in scenario 3. The results are shown for $\epsilon = 0.5$, three distances of the regression curves $d \in \{1, 1.25, 1.5\}$ and three different combinations of variances $(\sigma_1^2, \sigma_2^2) \in \{(0.25, 0.25), (0.25, 0.5), (0.5, 0.5)\}.$}
        \label{fig:Ex3_Power} 
\end{figure}

As previously observed in Section \ref{sec:simu:MA:2} only few type I error rates are non-zero and these exceptions are still much smaller than the nominal level of $\alpha = 0.05$, reaching a maximum of only $0.009$ (all values can be found in table S4 of the supplementary material). 
Hence, the analysis focuses on the power of the tests which is shown in Figure \ref{fig:Ex3_Power}. 
The loss of power resulting from model misspecification is not as large as in Section \ref{sec:simu:MA:2} but still present. Especially if one of the models is falsely specified to be an emax model but the other one specified correctly, we observe a notable loss of power not only compared to using the true model but also compared to using model averaging.
Moreover, this effect is increasing with increasing sample size.
In contrast to Section \ref{sec:simu:MA:2}, the power resulting from using model averaging is notably smaller than the one observed when using the true model. However, compared to two out of the three misspecifications, the loss in power is extensively reduced.
In conclusion, if the models are misspecified we observe either type I error inflation or a lack of power, both often of substantial extend, in all three scenarios. Using model averaging considerably reduces these problems, often leading to similar results as knowing and using the true underlying model.

\section{Case study} \label{sec:case}
We illustrate the proposed methodology through a case study analysing the equivalence of time-response curves (also known as exposure duration-response curves) using data which was originally published by \citet{Ghallab2021}. The study aims to investigate dietary effects onto the gene expression.
The dataset consists of two groups of mice which were fed with two different diets and then sacrificed at different time points. The first one is a high-fat or ``Western" diet (WD) while the other one is a standard diet (SD). 
As no data has been collected in the first 3 weeks, they are not included into our analysis. Consequentially, the beginning of the study ($t=0$) resembles week 3 of the actual experiment.
Data is then observed at $t=0, 3, 9, 15, 21, 27, 33, 39$ and $45$ for the Western diet and at $t=0, 3, 27, 33, 39$ and $45$ for the standard diet with sample sizes $5, 5, 5, 5, 5, 5, 5, 4, 8$
and $7, 5, 5, 7, 3, 5$, respectively. 
For each group the gene expression of 20733 genes is measured in terms of gene counts. For our analysis, we focus on the 1000 genes \citet{Ghallab2021} classified as especially interesting due to high activity. 
Although gene expression is measured as count data, it is treated as continuous due to the very high number of counts. The raw count data is preprocessed in terms of the gene count normalisation conducted by \citet{Ghallab2021} and subsequent $\log_2$-transformation of the normalised counts as suggested by \citet{Duda2022, Duda2023}.

Using this data, we aim to investigate the equivalence of the time-gene expressions curves between the two diets at a 5 \% significance level. From \citet{Ghallab2021} is is known that there are quite large differences between the diets, such that for the majority of the genes we expect not to conclude equivalence. However, precisely for this reason it is of interest for which genes equivalence can be concluded nevertheless. As we are interested in the results for each gene separately and are not aiming for a global conclusion, we do not adjust for multiple testing. 

As time-response studies are relatively rare, no specific time-response models have been developed. Hence, dose-response models are deployed for time-response relations as well. Methodological review studies \citep[e.g.][]{Kappenberg2023} do also not distinguish between dose-response and time-response studies. In addition, it seems intuitive that the effects of the high-fat diet accumulate with increasing time of consumption in a similar manner as the effects in dose response-studies accumulate with increasing dose. 

As outlined by \citet{Ghallab2021}, the dose-response relations vary across genes such that there is no single model which fits to all of them and, hence, model uncertainty is present. In addition, the models cannot be chosen manually due to the high number of genes. 
We introduce frequentist model averaging using AIC-based weights and the equivalence tests are performed using hybrid CI, i.e. by conducting Algorithm \ref{alg1}b. 
We deploy the set of candidate models suggested by \citet{Duda2022}, i.e. a linear model \eqref{eq:linear}, a quadratic model \eqref{eq:quadratic}, an emax model \eqref{eq:emax}, an exponential model \eqref{eq:exp}, a sigmoid emax model \eqref{eq:sigemax} and a beta model \eqref{eq:beta}. This set of candidate models can capture quite diverse effects, as it includes linear and nonlinear, increasing and decreasing, monotone and non-monotone as well as convex, concave and sigmoid curves.

The ranges of the response variables, i.e. the ranges of $\log_2$(normalised counts), are not comparable across different genes. Hence, different equivalence thresholds are needed for each of the genes. As such thresholds can not be chosen manually due to the high number of genes, we determine the thresholds as a percentile of the range of the response variable. For a gene $g \in \{1,...,1000\}$ that is 
$$
\epsilon_g = \Tilde{\epsilon} \left( \max_{l,i} (\hat{y}_{gli}) - \min_{l,i} (\hat{y}_{gli}) \right),
$$
where $\Tilde{\epsilon} \in (0,1)$ is the corresponding percentile and $\Tilde{\epsilon} = 0.2$ or 0.25 would be typical choices. Alternatively, one can proceed the other way around, calculate 
$$
\Tilde{u}_g = \frac{u_g}{\max_{l,i} (\hat{y}_{gli}) - \min_{l,i} (\hat{y}_{gli})}
$$ and directly compare $\Tilde{u}_g$ to $\Tilde{\epsilon}$, i.e. the decision rules $\epsilon_g > u_g$ and $\Tilde{\epsilon} > \Tilde{u}_g$ are equivalent. 

\begin{figure}
        \centering
        \begin{overpic}[scale = 0.795]{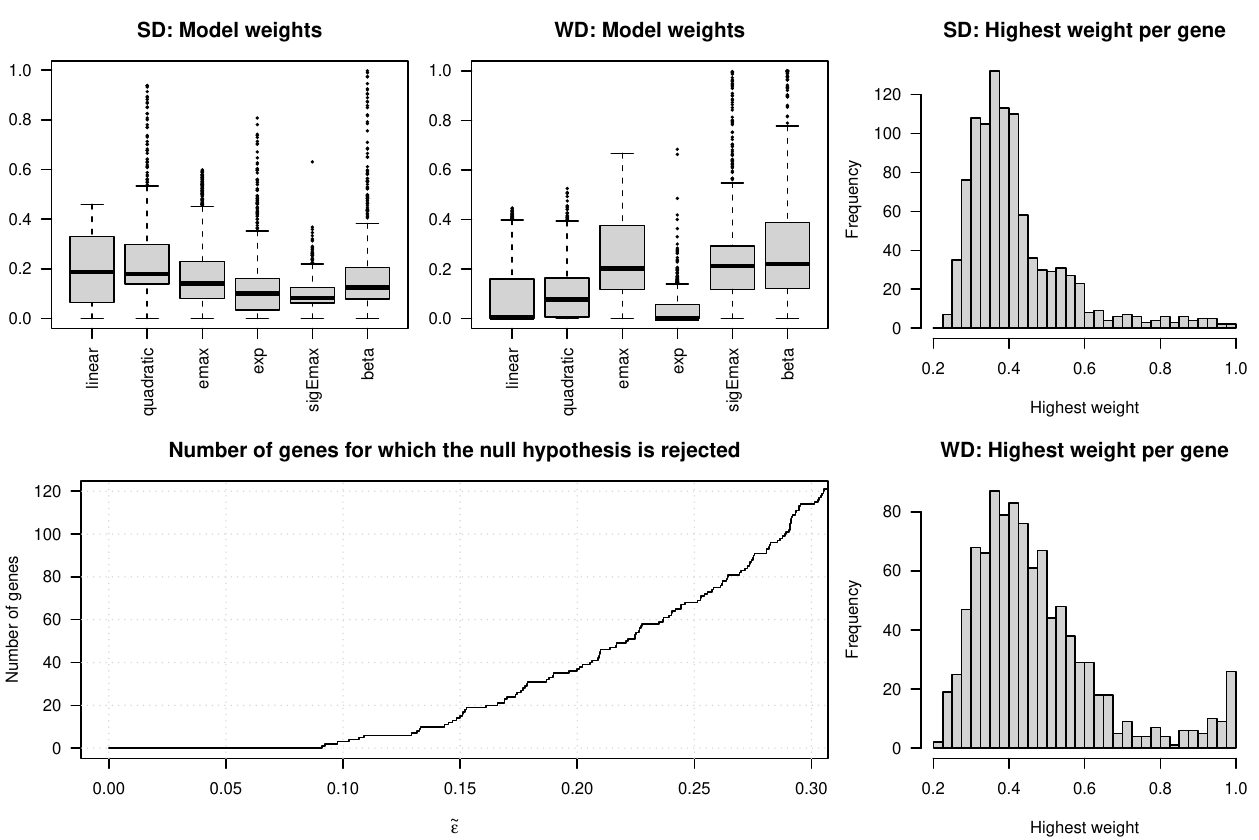} 
            \put(0,64){(a)}
            \put(69,64){(c)}
            \put(33.5,64){(b)}
            \put(69,30.5){(d)}
            \put(0,30.5){(e)}
        \end{overpic}
        \caption{Subfigures (a) and (b) show boxplots of the model weights for each of the two diets. Subfigures (c) and (d) show histograms of the highest model weight per gene for both genes. Subfigure (e) shows the number of genes for which equivalence between the time-gene expression curves of the two diets can be concluded in dependence of the equivalence threshold $\Tilde{\epsilon}$.}
        \label{fig:case_weights} 
\end{figure}
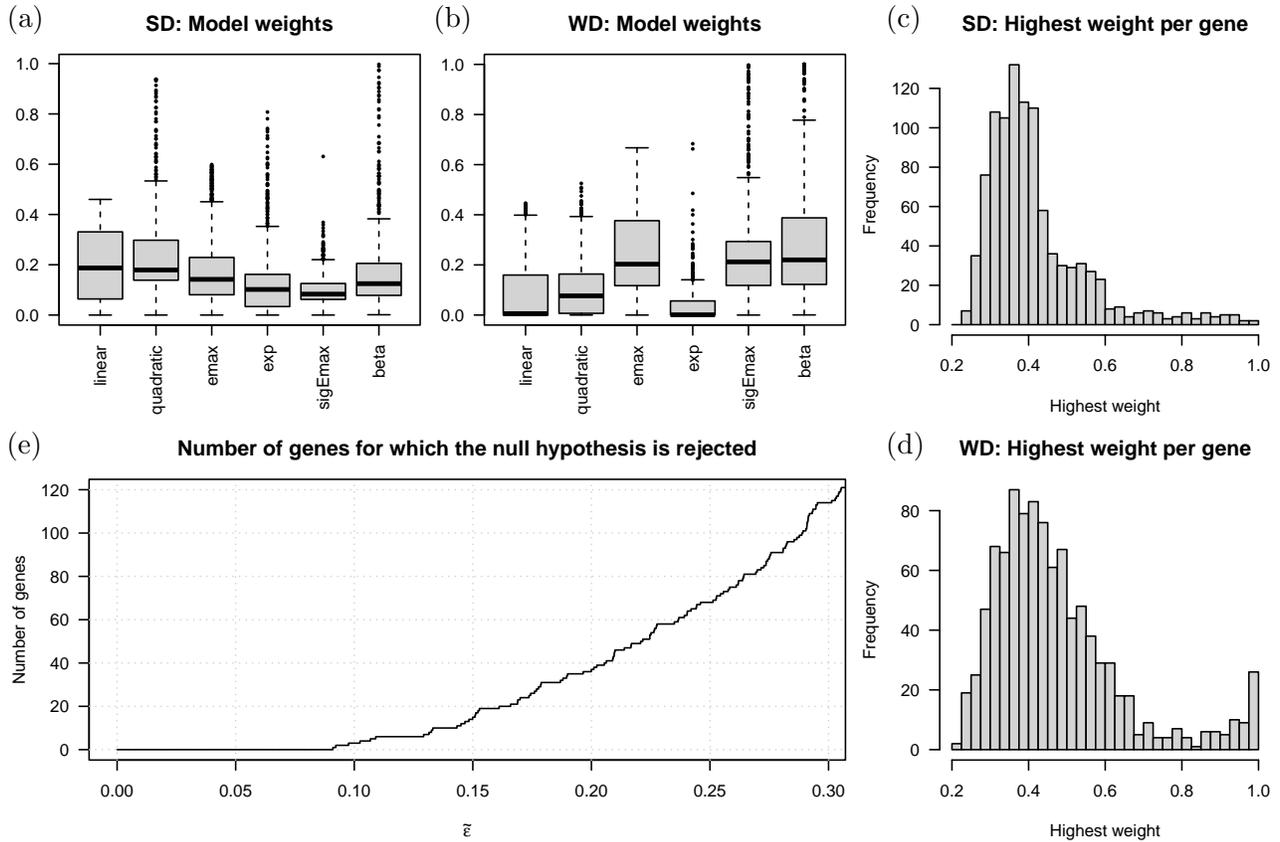

Subfigures (a) and (b) of Figure \ref{fig:case_weights} show boxplots of the model weights for both diets. It can be observed that less complex models, i.e. the linear and quadratic model, have higher weights for the standard diet compared to the Western diet. In contrast, for the two most complex models, i.e. the beta and sigEmax model, the opposite can be observed: they have higher weights for the Western diet compared to the standard diet.
In addition, Subfigures (c) and (d) of Figure \ref{fig:case_weights} show histograms of the highest model weight per gene. It can be observed that for the Western diet the model weights tend to be larger compared to the standard diet. In addition, for the Western diet there are notably many genes for which the highest model weight is very close to 1, i.e. the averaged model consists nearly fully of only one of the candidate models.

Subfigure (e) of Figure \ref{fig:case_weights} shows the number of genes for which equivalence of the time-gene expression curves of both diets can be concluded, i.e. the number of genes for which H$_0$ can be rejected depending on the choice of $\Tilde{\epsilon}$. For very small choices of $\Tilde{\epsilon}$ (e.g. 0.05 or 0.075)  equivalence cannot be concluded for any gene and for $\Tilde{\epsilon} = 0.1$ only three genes would be assessed as equivalent. 
For more typical choices of $\Tilde{\epsilon}$ being 0.2, 0.25 or 0.3, equivalence could be concluded for 37, 68 and 114 genes, respectively. With further increasing $\Tilde{\epsilon}$ the number of rejections also further increases and approaches 1000. However, this is not shown for $\Tilde{\epsilon} > 0.3$ as performing an equivalence test with a threshold larger than 30 \% of the range of the response variable might not have practical relevance. 

\begin{figure}
        \centering
        \includegraphics[scale = 0.795]{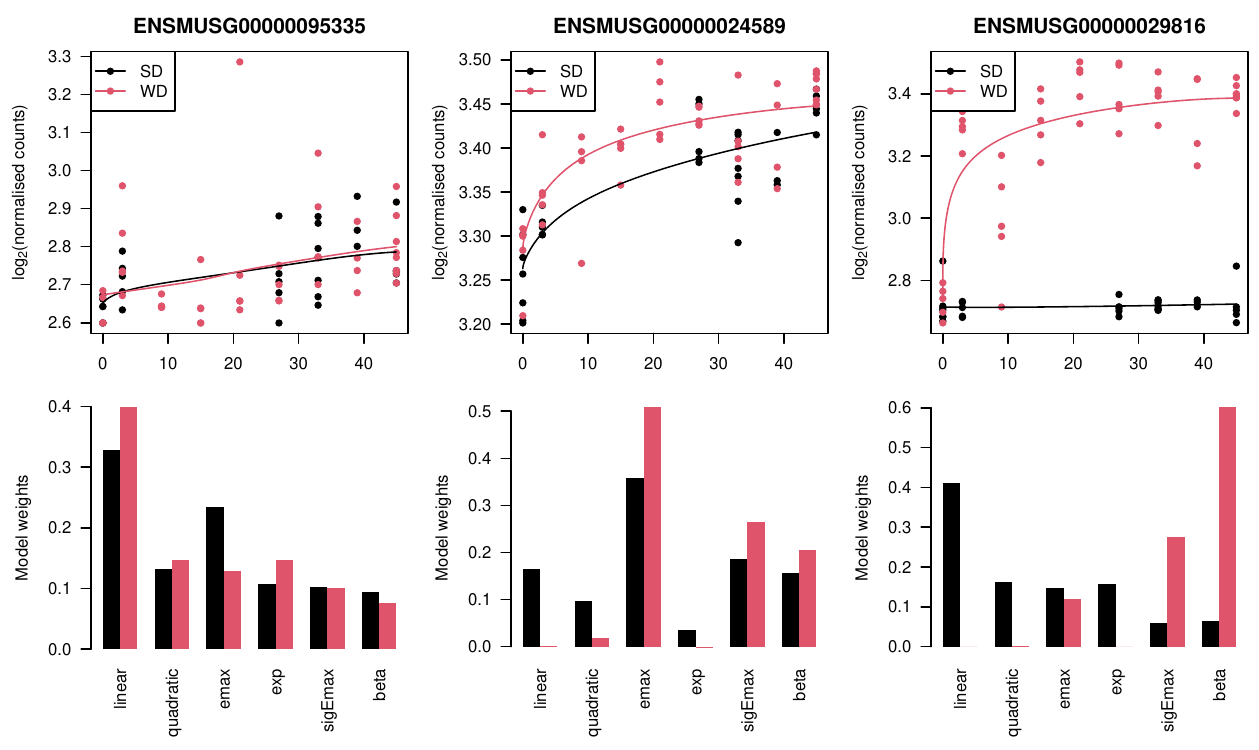} 
        \caption{Results for three exemplary genes. The first row of figures shows the data for both diets as well as the fitted models. The second row of figures shows the corresponding model weights.}
        \label{fig:case_curves} 
\end{figure}

Figure \ref{fig:case_curves} shows the results for three exemplary genes. For ENSMUSG00000095335 it can be observed that both time-response curves are extremely close to each other and that the maximum absolute deviation of the curves is quite small. This leads to $\Tilde{u}=0.098$. Regarding the model weights it can be observed that both time-response curves consist essentially of the same models. 
For ENSMUSG00000024589 we observe that both time-response curves have a similar shape both being emax-like, although their model weights are not as similar as before. However, their distance is larger than for ENSMUSG00000095335 which leads to $\Tilde{u} \approx 0.275$. Hence, the curves are not equivalent for typical choices of $\Tilde{\epsilon}$ being e.g. 0.2 or 0.25 but only for a more liberal choice of $\Tilde{\epsilon}= 0.3$. For the last example ENSMUSG00000029816 we observe that the two curves are completely different with regard to both, shape and location. For the standard diet an almost constant curve is present while for the Western diet a typical emax shape is observable. This is also reflected by the model weights where models which have high weights for one curve, have small ones for the other one and vise versa, the only exempt to this is the emax model which has a medium large weight for both of the groups. Due to the large maximum absolute deviation between the curves given by $\hat d=0.847$, similarity cannot be concluded for any reasonable equivalence threshold.  

\section{Conclusion} 
\label{sec:conclusion}
In this paper, we introduced a new approach for
model-based equivalence testing which can also be applied in the presence of model uncertainty -- a problem which is usually faced in practical applications. 
Our approach is based on a flexible model averaging method which relies on information criteria and a testing procedure which makes use of the duality of tests and confidence intervals rather than simulating the distribution under the null hypothesis, providing a numerically stable procedure. 
Moreover, our approach leads to additional interpretability due to the provided confidence intervals while retaining the asymptotic validity and a similar performance in finite samples as the bootstrap based test proposed by \citet{Dette2018}. 

Precisely, we investigated the finite sample properties of the proposed method by reanalysing the simulation study of \citet{Dette2018} and observed similar results for the CI-based test compared to their test. 
In the presence of model uncertainty, model misspecification frequently led to either type I error inflation or a lack of power, both often of substantial extend. In contrast, our approach considerably reduced these problems and in many cases even achieved similar results as knowing and using the true underlying model. 
The presented case study outlines the practical usefulness of the proposed method based on a large data application where choosing the models manually would be time-consuming and could easily lead to many model misspecifications.  
Hence, introducing model averaging here is essential to test for the equivalence of time-gene expression curves for such large numbers of genes, typically occurring in practice.

Future possible research includes extending the presented method for other model averaging techniques, e.g. cross validation-based model averaging. In addition, transferring this approach to other model classes (e.g. survival models) as well as to multidimensional responses, i.e. multiple endpoints, merits further exploration.

\section*{Software and data availability}
Software in the form of R code  available at \url{https://github.com/Niklas191/equivalence_tests_with_model_averaging.git}.
The case study data set is publicly available at the SRA database with reference number \hyperlink{https://www.ncbi.nlm.nih.gov/sra/PRJNA953810}{PRJNA953810}.

\section*{Funding}
This work has been supported by the Research Training Group "Biostatistical Methods for High-Dimensional Data in Toxicology`` (RTG 2624, P7) funded by the Deutsche Forschungsgemeinschaft (DFG, German Research Foundation, Project Number 427806116).

\section*{Competing interests}
The authors declare no competing interests.


\begin{thebibliography}{}

\bibitem[Aitchison, 1964]{Aitchison1964}
Aitchison, J. (1964).
\newblock Confidence-region tests.
\newblock {\em Journal of the Royal Statistical Society: Series B
  (Methodological)}, 26(3):462--476.

\bibitem[Akaike, 1974]{AIC}
Akaike, H. (1974).
\newblock A new look at the statistical model identification.
\newblock {\em IEEE Transactions on Automatic Control}, 19(6):716 -- 723.

\bibitem[Aoki et~al., 2017]{Aoki2017}
Aoki, Y., R{\"o}shammar, D., Hamr{\'e}n, B., and Hooker, A.~C. (2017).
\newblock Model selection and averaging of nonlinear mixed-effect models for
  robust phase iii dose selection.
\newblock {\em Journal of pharmacokinetics and pharmacodynamics}, 44:581--597.

\bibitem[Bastian et~al., 2024]{Bastian2024}
Bastian, P., Dette, H., Koletzko, L., and M\"ollenhoff, K. (2024).
\newblock Comparing regression curves -- an $l^1$-point of view.
\newblock {\em Annals of the Institute of Statistical Mathematics},
  76(1):159--183.

\bibitem[Bornkamp, 2015]{Bornkamp2015}
Bornkamp, B. (2015).
\newblock Viewpoint: model selection uncertainty, pre-specification, and model
  averaging.
\newblock {\em Pharmaceutical Statistics}, 14(2):79--81.

\bibitem[Bornkamp et~al., 2009]{Bornkamp2009}
Bornkamp, B., Pinheiro, J., and Bretz, F. (2009).
\newblock Mcpmod: An r package for the design and analysis of dose-finding
  studies.
\newblock {\em Journal of Statistical Software}, 29(7):1--23.

\bibitem[Breiman, 1996]{Breiman1996}
Breiman, L. (1996).
\newblock Heuristics of instability and stabilization in model selection.
\newblock {\em Annals of Statistics}, 24:2350--2383.

\bibitem[Bretz et~al., 2018]{Bretz2018}
Bretz, F., M\"ollenhoff, K., Dette, H., Liu, W., and Trampisch, M. (2018).
\newblock Assessing the similarity of dose response and target doses in two
  non-overlapping subgroups.
\newblock {\em Statistics in Medicine}, 37(5):722--738.

\bibitem[Bretz et~al., 2005]{Bretz2005}
Bretz, F., Pinheiro, J.~C., and Branson, M. (2005).
\newblock Combining multiple comparisons and modeling techniques in
  dose-response studies.
\newblock {\em Biometrics}, 61(3):738--748.

\bibitem[Buatois et~al., 2018]{Buatois2018}
Buatois, S., Ueckert, S., Frey, N., Retout, S., and Mentr{\'e}, F. (2018).
\newblock Comparison of model averaging and model selection in dose finding
  trials analyzed by nonlinear mixed effect models.
\newblock {\em The AAPS journal}, 20:1--9.

\bibitem[Buckland et~al., 1997]{Buckland1997}
Buckland, S.~T., Burnham, K.~P., and Augustin, N.~H. (1997).
\newblock Model selection: An integral part of inference.
\newblock {\em Biometrics}, 53(2):603--618.

\bibitem[Claeskens and Hjort, 2008]{Claeskens_2008}
Claeskens, G. and Hjort, N.~L. (2008).
\newblock {\em Model Selection and Model Averaging}.
\newblock Cambridge Series in Statistical and Probabilistic Mathematics.
  Cambridge University Press.

\bibitem[Dennis et~al., 2019]{Dennis2019}
Dennis, B., Ponciano, J.~M., Taper, M.~L., and Lele, S.~R. (2019).
\newblock Errors in statistical inference under model misspecification:
  Evidence, hypothesis testing, and aic.
\newblock {\em Frontiers in Ecology and Evolution}, 7.

\bibitem[Dette et~al., 2018]{Dette2018}
Dette, H., M\"ollenhoff, K., Volgushev, S., and Bretz, F. (2018).
\newblock Equivalence of regression curves.
\newblock {\em Journal of the American Statistical Association},
  113(522):711--729.

\bibitem[Duda et~al., 2023]{Duda2023}
Duda, J., Drenda, C., K\"astel, H., Rahnenf\"uhrer, J., and Kappenberg, F.
  (2023).
\newblock Benefit of using interaction effects for the analysis of
  high-dimensional time-response or dose-response data for two-group
  comparisons.
\newblock {\em Scientific Reports}, 13:20804.

\bibitem[Duda et~al., 2022]{Duda2022}
Duda, J.~C., Kappenberg, F., and Rahnenf\"uhrer, J. (2022).
\newblock Model selection characteristics when using mcp-mod for dose-response
  gene expression data.
\newblock {\em Biometrical Journal}, 64(5):883--897.

\bibitem[Efron and Tibshirani, 1994]{Efron1994}
Efron, B. and Tibshirani, R.~J. (1994).
\newblock {\em An Introduction to the Bootstrap}.
\newblock Monographs on Statistics \& Applied Probability. Chapman and Hall,
  New York.

\bibitem[Fletcher, 2018]{Fletcher_2018}
Fletcher, D. (2018).
\newblock {\em Model Averaging}.
\newblock Springer Briefs in Statistics. Springer.

\bibitem[Ghallab et~al., 2021]{Ghallab2021}
Ghallab, A., Myllys, M., Friebel, A., Duda, J., Edlund, K., Halilbasic, E.,
  Vucur, M., Hobloss, Z., Brackhagen, L., Begher-Tibbe, B., Hassan, R., Burke,
  M., Genc, E., Frohwein, L.~J., Hofmann, U., Holland, C.~H., Gonz\'alez, D.,
  Keller, M., Seddek, A.-l., Abbas, T., Mohammed, E. S.~I., Teufel, A., Itzel,
  T., Metzler, S., Marchan, R., Cadenas, C., Watzl, C., Nitsche, M.~A.,
  Kappenberg, F., Luedde, T., Longerich, T., Rahnenf\"uhrer, J., Hoehme, S.,
  Trauner, M., and Hengstler, J.~G. (2021).
\newblock Spatio-temporal multiscale analysis of western diet-fed mice reveals
  a translationally relevant sequence of events during nafld progression.
\newblock {\em Cells}, 10(10):1--29.

\bibitem[Gsteiger et~al., 2011]{Gsteiger2011}
Gsteiger, S., Bretz, F., and Liu, W. (2011).
\newblock Simultaneous confidence bands for nonlinear regression models with
  application to population pharmacokinetic analyses.
\newblock {\em Journal of Biopharmaceutical Statistics}, 21(4):708--725.

\bibitem[Guhl et~al., 2022]{Guhl2022}
Guhl, M., Mercier, F., Hofmann, C., Sharan, S., Donnelly, M., Feng, K., Sun,
  W., Sun, G., Grosser, S., Zhao, L., Fang, L., Mentr\'e, F., Comets, E., and
  Bertrand, J. (2022).
\newblock Impact of model misspecification on model-based tests in pk studies
  with parallel design: real case and simulation studies.
\newblock {\em Journal of Pharmacokinetics and Pharmacodynamics},
  49(5):557--577.

\bibitem[Hagemann et~al., 2024]{Hagemann2024}
Hagemann, N., Marra, G., Bretz, F., and M\"ollenhoff, K. (2024).
\newblock Testing for similarity of multivariate mixed outcomes using
  generalised joint regression models with application to efficacy-toxicity
  responses.
\newblock {\em arXiv: 2401.05817 [stat.ME]}.

\bibitem[Hauschke et~al., 2007]{Hauschke2007}
Hauschke, D., Steinijans, V., and Pigeot, I. (2007).
\newblock {\em Bioequivalence Studies in Drug Development}.
\newblock John Wiley \& Sons, Ltd.

\bibitem[Jhee et~al., 2004]{Jhee2004}
Jhee, S.~S., Lyness, W.~H., Rojas, P.~B., Leibowitz, M.~T., Zarotsky, V., and
  Jacobsen, L.~V. (2004).
\newblock Similarity of insulin detemir pharmacokinetics, safety, and
  tolerability profiles in healthy caucasian and japanese american subjects.
\newblock {\em The Journal of Clinical Pharmacology}, 44(3):258--264.

\bibitem[Kappenberg et~al., 2023]{Kappenberg2023}
Kappenberg, F., Duda, J., Sch\"urmeyer, L., G\"ul, O., Brecklinghaus, T.,
  Hengstler, J., Schorning, K., and Rahnenf\"uhrer, J. (2023).
\newblock Guidance for statistical design and analysis of toxicological
  dose-response experiments, based on a comprehensive literature review.
\newblock {\em Archives of Toxicology}, 97(10):2741 -- 2761.

\bibitem[Lakens, 2017]{Lakens2017}
Lakens, D. (2017).
\newblock Equivalence tests: A practical primer for t tests, correlations, and
  meta-analyses.
\newblock {\em Social Psychological and Personality Science}, 8(4):355--362.

\bibitem[Leeb and P{\"o}tscher, 2005]{Leeb2005}
Leeb, H. and P{\"o}tscher, B.~M. (2005).
\newblock Model selection and inference: Facts and fiction.
\newblock {\em Econometric Theory}, 21(1):21--59.

\bibitem[Leeb and P{\"o}tscher, 2008]{Leeb2008}
Leeb, H. and P{\"o}tscher, B.~M. (2008).
\newblock Can one estimate the unconditional distribution of
  post-model-selection estimators?
\newblock {\em Econometric Theory}, 24(2):338--376.

\bibitem[Ley and Steel, 2009]{Ley2009}
Ley, E. and Steel, M.~F. (2009).
\newblock On the effect of prior assumptions in bayesian model averaging with
  applications to growth regression.
\newblock {\em Journal of Applied Econometrics}, 24(4):651--674.

\bibitem[Liu et~al., 2009]{Liu2009}
Liu, W., Bretz, F., Hayter, A.~J., and Wynn, H.~P. (2009).
\newblock Assessing nonsuperiority, noninferiority, or equivalence when
  comparing two regression models over a restricted covariate region.
\newblock {\em Biometrics}, 65(4):1279--1287.

\bibitem[M\"ollenhoff et~al., 2024]{Moellenhoff2024}
M\"ollenhoff, K., Binder, N., and Dette, H. (2024).
\newblock Testing similarity of parametric competing risks models for
  identifying potentially similar pathways in healthcare.
\newblock {\em arXiv: 2401.04490 [stat.ME]}.

\bibitem[M\"ollenhoff et~al., 2020]{Moellenhoff2020}
M\"ollenhoff, K., Bretz, F., and Dette, H. (2020).
\newblock Equivalence of regression curves sharing common parameters.
\newblock {\em Biometrics}, 76(2):518--529.

\bibitem[M\"ollenhoff et~al., 2021]{Moellenhoff2021}
M\"ollenhoff, K., Dette, H., and Bretz, F. (2021).
\newblock {Testing for similarity of binary efficacy-toxicity responses}.
\newblock {\em Biostatistics}, 23(3):949--966.

\bibitem[M\"ollenhoff et~al., 2018]{Moellenhoff2018}
M\"ollenhoff, K., Dette, H., Kotzagiorgis, E., Volgushev, S., and Collignon, O.
  (2018).
\newblock Regulatory assessment of drug dissolution profiles comparability via
  maximum deviation.
\newblock {\em Statistics in Medicine}, 37(20):2968--2981.

\bibitem[M\"ollenhoff et~al., 2022]{moellenhoff2022}
M\"ollenhoff, K., Loingeville, F., Bertrand, J., Nguyen, T.~T., Sharan, S.,
  Zhao, L., Fang, L., Sun, G., Grosser, S., Mentr\'{e}, F., and Dette, H.
  (2022).
\newblock {Efficient model-based bioequivalence testing}.
\newblock {\em Biostatistics}, 23(1):314--327.

\bibitem[Otto et~al., 2008]{Otto2008}
Otto, C., Fuchs, I., Altmann, H., Klewer, M., Walter, A., Prelle, K., Vonk, R.,
  and Fritzemeier, K.-H. (2008).
\newblock {Comparative Analysis of the Uterine and Mammary Gland Effects of
  Drospirenone and Medroxyprogesterone Acetate}.
\newblock {\em Endocrinology}, 149(8):3952--3959.

\bibitem[Pinheiro et~al., 2014]{Pinheiro2014}
Pinheiro, J., Bornkamp, B., Glimm, E., and Bretz, F. (2014).
\newblock Model-based dose finding under model uncertainty using general
  parametric models.
\newblock {\em Statistics in Medicine}, 33(10):1646--1661.

\bibitem[Pinheiro et~al., 2006]{Pinheiro2006}
Pinheiro, J.~C., Bretz, F., and Branson, M. (2006).
\newblock {\em Analysis of Dose--Response Studies--Modeling Approaches}.
\newblock Dose Finding in Drug Development. Springer, New York.

\bibitem[Price et~al., 2011]{Price2011}
Price, M.~J., Welton, N.~J., Briggs, A.~H., and Ades, A. (2011).
\newblock Model averaging in the presence of structural uncertainty about
  treatment effects: influence on treatment decision and expected value of
  information.
\newblock {\em Value in Health}, 14(2):205--218.

\bibitem[Schorning et~al., 2016]{Schorning2016}
Schorning, K., Bornkamp, B., Bretz, F., and Dette, H. (2016).
\newblock Model selection versus model averaging in dose finding studies.
\newblock {\em Statistics in Medicine}, 35(22):4021--4040.

\bibitem[Schuirmann, 1987]{Schuirmann1987}
Schuirmann, D.~J. (1987).
\newblock A comparison of the two one-sided tests procedure and the power
  approach for assessing the equivalence of average bioavailability.
\newblock {\em Journal of Pharmacokinetics and Biopharmaceutics}, 15:657--680.

\bibitem[Schwarz, 1978]{BIC}
Schwarz, G. (1978).
\newblock Estimating the dimension of a model.
\newblock {\em The Annals of Statistics}, 6(2):461--464.

\bibitem[Wasserman, 2000]{Wasserman2000}
Wasserman, L. (2000).
\newblock Bayesian model selection and model averaging.
\newblock {\em Journal of Mathematical Psychology}, 44(1):92--107.

\end{thebibliography}

\end{document}